\documentclass[12pt]{iopart}
\usepackage{graphicx}
\usepackage{colordvi}
\usepackage{colortbl}
\usepackage{ulem}
\usepackage{dcolumn}% Align table columns on decimal point
\usepackage{bm}% bold math

\newcommand{\Rvec}{{\bf R}}
\newcommand{\qvec}{{\bf q}}
\newcommand{\Qvec}{{\bf Q}}
\newcommand{\kvec}{{\bf k}}

\begin{document}
\title{Dynamical charge and spin density wave scattering in cuprate superconductors}
\author{G. Seibold}                                                           
\address{Institut f\"ur Physik, BTU Cottbus, PBox 101344, 03013 Cottbus, 
Germany}                                                                       
\author{M. Grilli, J. Lorenzana}                                               
\address{ISC-CNR and Dipartimento di Fisica,                             
Universit\`a di Roma ``La Sapienza'', P.le Aldo Moro 5, I-00185 Roma, Italy}

\begin{abstract}
We show that a variety of spectral features in high-T$_c$ cuprates
can be understood from the coupling of charge carriers to 
some kind of dynamical order which we exemplify in terms
of fluctuating charge and spin density waves. Two theoretical
models are investigated which capture different aspects
of such dynamical scattering. The first approach leaves
the ground state in the disordered phase but couples
the electrons to bosonic degrees of freedom, corresponding
to the quasi singular scattering associated with 
the closeness to an ordered phase. The second, more phenomological
approach starts from the construction of a frequency
dependent order parameter which vanishes for small 
energies. Both theories capture scanning tunneling microscopy
and angle-resoved photoemission experiments which suggest the
protection of quasiparticles close to the Fermi energy but
the manifestation of long-range order at higher frequencies.
\end{abstract}
\maketitle

\section{Introduction}
Two major lines of thought are presently debated for the understanding of the
high-temperature superconducting cuprates. 
On the one hand \cite{anderson1,anderson2,LNW}
these systems are seen as doped Mott insulators, where strong electron-electron 
correlations only play the major role. In this case at low doping (the underdoped
region), short-range antiferromagnetic correlations spoil the metallic Fermi liquid
(FL) phase producing singlet pairs, which give rise to a pseudogap below a temperature $T^*$
and eventually condensing into a superconducting state below $T_c<T^*$. Upon increasing doping, correlations
progressively loose strength and $T^*$ merges into the $T_c$ line decreasing in the overdoped region. 
The second point of view \cite{cast95,chubukov,varma,reviewQCP}
is that a more or less hidden electronic order is present in underdoped cuprates.
For sure correlations favor the occurrence of this order, but normal-state anomalies and
high-temperature superconductivity essentially stem from the presence of this order:
The proximity to an instability (particularly if it is a second-order ``critical line'' ending 
at zero temperature into a quantum critical point)
marking the onset of order naturally brings along abundant fluctuations and leads to
strongly temperature- and doping-dependent features, which accounts for the non-FL properties and for a 
strong pairing interaction. These two distinct
points of view particularly face themselves in the so-called ``glue issue'' \cite{glue1,glue2}: 
what is the source of strong scattering/pairing between the charge carriers leading to high-T$_c$? 
In the case of a doped Mott insulator,
one naturally expects a nearly instantaneous magnetic coupling $J$ to
play the role of glue in forming the 
singlets, while for the ``quantum critical'' scenario the
order-parameter fluctuations provide a glue 
mechanism, which is inherently retarded due to the slow dynamics of
bosonic critical fluctuations.  
In this situation, we find that the two debated issues in the
cuprates, namely  to identify the source of electronic 
scattering and to identify the phase (if any) competing or coexisting
with pairing in the underdoped region,  
are the two faces of the same medal. 

From the above discussion it should be clear that to detect and to
characterize any  
boson-like excitation coupled to the quasiparticles (QPs) is a major
issue. This is precisely the scope of our 
paper, where we focus on the spectroscopic effects of order-parameter
fluctuations. We will show that their 
dynamical character can make them rather elusive at low energy, while
they leave clear signatures at 
higher energies.  Moreover, the specific wavevector dependence of
these excitations, 
as inferred from experiments, indicates that 
these retarded bosonic excitations are due to dynamical
charge and spin ordering fluctuations. 
This implies that a locally ordered dynamical state is the natural
candidate as the competing 
phase of underdoped cuprates. Far from being alternative to charge
ordering fluctuations, we also 
believe \cite{CGDMH} that spin fluctuations are also relevant since
they are sustained and  supported up to large dopings by the occurrence of fluctuating
charge-depleted regions.  
In the next Section we will briefly overview previous results within
this fluctuating order scenario. 
In Sect. 3 we will present the general theoretical framework, which
will find two distinct 
phenomenological realizations in the subsequent Sects.4 and 5. Our
concluding remarks are 
reported in Sect. 6.

\section{Charge Ordering: a brief overview}
\label{sec:charge-order-brief}
Since the discovery of high-temperature superconductors by Bednorz and
M\"uller \cite{bm} numerous experiments have evidenced the existence
of electronic inhomogeneities in these compounds (cf. e.g. Ref. \cite{ps}).
While early on these inhomogeneities where believed to be predominantly
due to material imperfections like disorder induced by the dopant ions,
it was subsequently realized that the strongly correlated character
of the cuprate superconductors and thus the electronic subsystem
itself  favors the formation of the inhomogeneities.
According to the theoretical analyis in Ref. \cite{cast95} the
reduction of the kinetic energy of the doped charge carriers caused by
the strong correlations together with a short range attractive force,
provided e.g. by electron-lattice interactions, gives rise to
a phase separation instability. On the other hand, the long-range
repulsive Coulomb interaction will spoil the associated zero-momentum
instability in the charge sector and instead shift the wave-vector
of the ordering transition to finite values which thus corresponds
to an incommensurate charge ordering (CO). This is the so-called frustrated phase
separation\cite{FPS}.
  
Alternatively, Hartree-Fock investigations of Hubbard (and tJ)-type 
hamiltonians
\cite{zaa89,mac89,hsch90,pol89} have suggested early on that these
models favor solutions with a combined charge- and spin-density wave.
These solutions, which have been confirmed later by more sophisticated
numerical methods (cf. e.g. \cite{WS,fle01b,HELLBERG}), are characterized by
one-dimensional hole-enriched domain walls where the 
antiferromagnetic (AF) order changes sign. 

The existence of such textures in lanthanum cuprates, codoped
with Nd, was confirmed by elastic neutron scattering experiments
\cite{tra95,tra96,tra97}. These so called 'stripes' have
also been found in other cuprate materials, codoped with Ba \cite{fujita02}
or Eu \cite{klauss00} and the associated CO ordering has been 
explicitely established by soft resonant x-ray scattering
\cite{abb05, fink09}. 

How generic are these charge- and spin density waves in the family
of high-T$_c$ materials? First, it is interesting to observe that
the non-codoped lanthanum cuprates show strong similarities in
the spin channel to their codoped counterparts. This includes not only
the doping dependence of the low energy incommensurability \cite{yam98},
but also the spectrum of high energy magnon excitations which
shows the same 'hour glass' shape in La$_{2-x}$Ba$_{x}$CuO$_4$ (LBCO) 
\cite{tra04} 
and La$_{2-x}$Sr$_{x}$CuO$_4$ (LSCO) \cite{chr04}.
These features can be well described on the basis of striped
ground states \cite{lor02,sei05,sei06} which also can account
for the doping dependence of mid-infrared excitation in
LSCO \cite{lor03}. 
The 'hour glass' magnetic spectrum is now also well established in
YBCO superconductors \cite{hay04} where in the strongly underdoped
regime even a static incommensurate spin response has been 
observed \cite{hinkov08}.

While neutron scattering provides a clear picture in the spin channel
a similar tool for the charge channel does not exist. Electron energy
loss spectroscopy or resonant inelastic x-ray scattering, although
potentially being  such tool, do not have yet the
resolution to investigate the dynamical charge response with a similar
accuracy as the one we have for spin fluctuations. Thus we are rather ``blind'' 
with respect to charge fluctuations.   

One can hope to see evidence of charge and spin order in the one
particle spectral properties accessible through angle resolved
photoemission spectroscopy (ARPES) and scanning tunneling spectroscopy. 
However, direct evidence for  charge and spin order is quite elusive
at low energies and in this paper we argue that the reason for this 
'invisibility' may be due to the dynamical nature of the scattering.
We will discuss two different flavors of dynamical charge or spin
ordering. Both of them have 
``protected'' low-energy QPs and thus an untouched  Fermi surface (FS).
However, in the first version there is no long range order, while in the second
version long range order is present. Within the former 
scenario the system is characterized by long (but finite)
range spatial correlations which dynamics is associated with a
characteristic time scale $\tau \sim 1/\omega_0$. Thus electrons
with energy (measured with respect to the Fermi energy) 
$\omega < \omega_0$ will average over the fluctuations of the order
parameter and
therefore keep the QP properties of an ordinary homogeneous FL.
On the other hand, a snapshot on time scales $t < \tau$ (i.e. frequencies
$\omega > \omega_0$) will detect an almost ordered system and
we thus expect the features of a conventional ordered system to become
apparent in the spectral function at large enough energies. The idea 
of a low energy sector with protected quasiparticles and a high-energy
sector displaying some form of order is also closely related to a
recent model of FeAs based superconductors\cite{dai08}.

\begin{figure}[thb]
\includegraphics[angle=-90,width=8cm,clip=true]{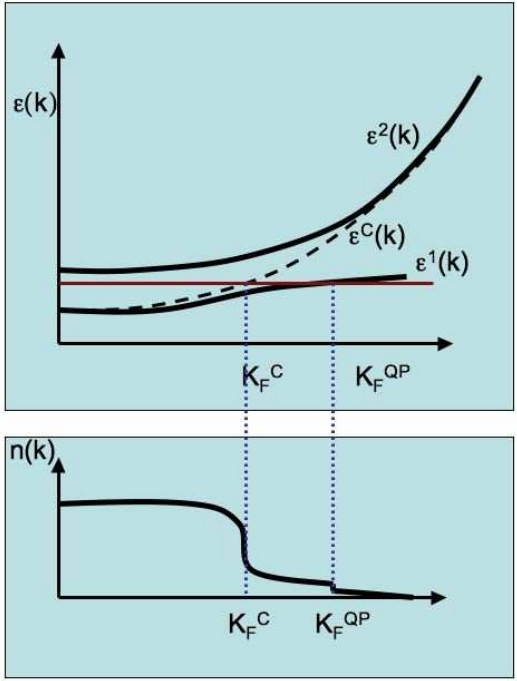}
\caption{Schematic view of a heavy-fermion system with a Kondo-like resonance 
arising at the Fermi energy E$_F$
from the mixing of a deep narrow $f$ level (not shown) and the conduction 
band (dashed line). Two QP bands, 
$\varepsilon^{1,2}(k)$, arise. The corresponding momentum distribution function $n(k)$ 
is shown below with a true Fermi momentum
$k_F^{QP}$ and a ``fictitious'' FS at $k_F^c$, where the conduction 
band (dashed line) crosses
the Fermi level (thin red line).
} 
\label{fig-HF}
\end{figure}

This idea of (incoherent) high-energy states carrying 
a specific momentum structure might seem strange, but
another example can clarify the concept of an incoherent part with a
strong momentum dependence which carries  physical information on
the short range physics. Lets consider the more standard issue of
large and small FSs in heavy fermions represented in
Fig.~\ref{fig-HF}. 
In heavy-fermions strongly correlated electrons in a  narrow
half-filled $f$ level hybridize with 
electrons in a conduction band and give rise to a Kondo resonance at the Fermi level formed by coherent
QP states. The width (and weight) of this QP band is usually quite small and sets the scale of the coherence
energy in these systems. Now consider the momentum distribution
function defined by
\begin{equation}
n_\kvec \equiv \int d\omega A(\kvec, \omega) f(\omega).
\label{nk}
\end{equation}
Here $f(\omega)$ is the Fermi function and the spectral density 
\begin{equation} 
A(\kvec,\omega)\equiv\frac{1}{\pi}ImG(\kvec,\omega)=\frac{1}{\pi}
\frac{\Sigma''(\kvec,\omega)}{\left(\omega-\Sigma'(\kvec,\omega) \right)^2+\Sigma''^2(\kvec,\omega)}
\end{equation}
is proportional to the imaginary part of the electron Green's function with real (imaginary) part
of the self-energy $\Sigma'$ $(\Sigma'')$.
One can see that $n_\kvec$ involves all the excitation energies and its features
might be dominated by the incoherent part of the spectrum if the QPs
have a minor weight. Indeed strictly speaking the true FS at zero temperature is given by the small jump
in the Fermi distribution function determining the Fermi momenta of the QPs at $\kvec_F^{QP}$
(see Fig. \ref{fig-HF}). 
This FS is large and satisfies the Luttinger theorem with a number of carriers including the
electrons in the $f$ level. This FS would naturally be determined by following the QP dispersion
[$\varepsilon^1(\kvec)$ in the upper panel of Fig. \ref{fig-HF}]. On the other hand
the shape of $n_\kvec$ is substantially determined by the (incoherent)
part of the spectral function, which
has strong weight at energies corresponding to both the $f$ level and  the conduction band. This latter
gives rise to a rather sharp decrease of $n_\kvec$ at a ``fictitious''
Fermi momentum $\kvec_F^c$, corresponding 
to the FS that the electrons in the conduction band would have in the
absence of mixing with the f-level. If the hybridization between the
$f$ level and the conduction band is turned off so that the
QP weight $z$ is driven to zero one  reaches
a situation in which the decrease at $\kvec_F^c$ becomes a discontinuity 
and the small jump at $\kvec_F^{QP}$ disappears.   It is clear that the 
sharp decrease of spectral weight
at $\kvec_F^c$ for finite hybridization has strong physical content for
an observer who ignores the underlying model. Thus in this example we
see the recurrent picture of low energy quasiparticles appearing due
to coherence effects and high energy incoherent excitations with a relevant momentum
structure which carry information on the short range physics.  

\begin{figure}[thb]
\includegraphics[width=8cm,clip=true]{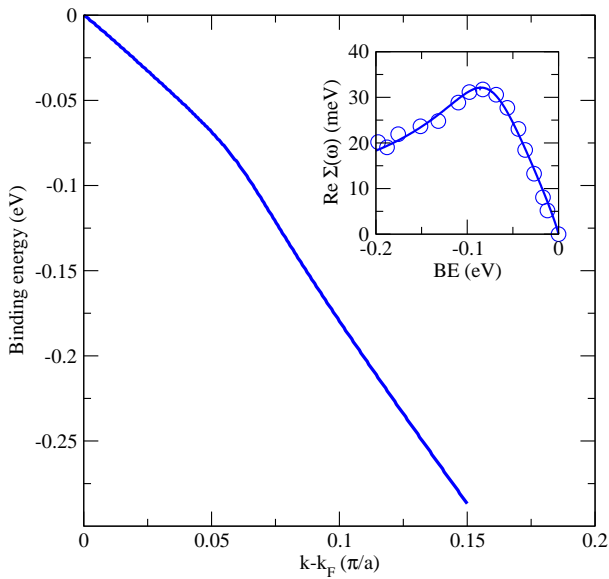}
\caption{Kink in the nodal electronic dispersion due to QPs coupled to CO fluctuations
with a Lorentzian frequency distribution centered at
$\omega_0=75$ meV and it has a width  $\gamma=40$ meV. 
The inset reports the
real part of the self-energy together with the experimental data
reported in Ref. \cite{GWEON}.} 
                        \label{fig-disp}
\end{figure}

In a series of papers \cite{sei01,grilli05,grilli09} we have worked 
out the scenario of Fermionic QP coupled to dynamical CO fluctuations
 with regard to several experimental observations. One first 
important remark is that all the conclusions drawn in the case
of QP coupled to spin excitations \cite{eschrig} might also be obtained
for the case of CO fluctuations (with, of course, suitable changes in the
energy scales and in the momentum dependencies). As an example we
report in Fig. \ref{fig-disp} the electronic dispersion along the BZ diagonal
obtained when the QPs are coupled to a  charge 
fluctuations with flat distribution of momenta and a Lorentzian 
distribution of frequencies
(see below, in Sect. IV). The typical frequency of the CO fluctuations
is $\omega_0=75$ meV, and the width of the distribution is $\gamma=40$ meV.
The bare electronic dispersion (with Fermi velocity $v_F=2/\pi$ eV
(in units of lattice spacing) is dressed by the real-part of the self-energy $\Sigma'$
reported in the inset together with the experimental data (open circles) 
of $\Sigma'$  from Ref. \cite{GWEON}. Ref. \cite{GWEON} also reported
a remarkable isotopic shift of the electronic dispersions, which had opposite
sign in different regions of the BZ. Although this effect is controversial
\cite{dessau,garcia}, we were able to show that an isotopic
dependence of the coherence length of CO dynamical fluctuations can reproduce
this effect and account for its strong momentum and doping dependence \cite{grilli05}.

Dynamical spin fluctuations, instead, would produce a similar effect, but with
a wrong momentum dependence. Another non trivial effect of dynamical
CO fluctuation is related to the dichotomy in the Fermi
surface of high-T$_c$ cuprates \cite{zho99,zho01}.
Especially these latter ARPES
studies on underdoped LSCO support our picture of dynamical order in  
the cuprates. On the one hand the momentum dependence of the  
low-energy part of the energy                                           
distribution curves was followed, thereby reconstructing the low-energy 
QP dispersion. In this way a large FS was found 
corresponding to the    
FL LDA band-structure and fulfilling the Luttinger            
requirement that the volume of the FS encircled the whole number of     
fermionic carriers $n=1-x$.                                             
On the other hand the FS was determined from the momentum distribution  
$n_\kvec$, obtained by integrating                                      
the spectral function over a broad energy window ($\sim 300$            
meV). Then the locus of momentum-space                                  
points where $n_\kvec$ displays a sharp decrease, marked a FS formed    
by two nearly parallel (weakly                                          
modulated) lines along the $k_x$ direction and crossing two similar
lines along the $k_y$ direction.
This crossed FS would naturally arise in a system with
one-dimensional stripes along the $y$ and $x$ 
directions. Thus the coexistence  of stripe-like spectral 
features at  large energies with a 'protected' FS at
low energies due to coherence effects exactly corresponds to the scenario of fluctuating
order sketched above.

Further even more direct evidence for dynamical CO in cuprate
superconductors is provided by scanning tunneling microscopy (STM)
experiments. STM investigations performed on bismuthate and 
oxychloride superconductors see a complex modulation
of the local density of states (LDOS) both in the superconducting (SC)      
state \cite{hoff02,how03,elroy05,hashi06,hana07,wise08} and above $T_c$     
\cite{wise08,versh04,hana04}. In both cases one observes peaks in           
the Fourier transform of the                                                
real space LDOS at wave-vectors $Q=2\pi/(4a_0)...2\pi/(5a_0)$ suggestive of 
checkerboard or stripe charge order. However, the debate is about the       
question whether these peaks are non-dispersive in energy (and thus signature 
of 'real' charge order) or follow a bias-dependent dispersion 
due to QP interference. 
In the latter case the spatial LDOS variations can be understood                
from the so-called octet model \cite{hoff02,wang03}                           
which attributes the modulations to the                                       
elastic scattering between the high density regions of the                    
Bogoljubov 'bananas' in the superconducting state.                            
Recent STM investigations \cite{kohsaka08,alldredge08,kohsaka07} may resolve this 
apparent conflict since they suggest that both, dispersive and non-dispersive  
scattering originates from different regions in momentum and energy space.    
The states in the nodal region                                                
which are well defined in k-space and undergo a transition                    
to a $d$-wave SC state below $T_c$ are then responsible for the low energy      
QP interference structure of the LDOS, whereas the ill-defined k-space 'quasiparticle'
states in the antinodal regions
are responsible for the non-dispersive CO above some energy
scale $\omega_0$. 
A phenomenological model of dynamical CO (but in the presence of long-range
order also) was recently adopted \cite{sei09} to describe this dichotomic 
behavior
observed in STM experiments. In particular, assuming a specific 
frequency-dependent
CO order parameter producing a marginal-FL type self-energy
\cite{varma91} we were able to reconcile                                  
the simultaneous existence of low energy Bogoljubov quasiparticles            
and high energy electronic order. Moreover the theory also accounts for the
CO specific  contrast reversal in the STM spectra between positive and
negative bias \cite{ma08} where the energy scale for the modulation 
of the LDOS is essentially determined by the pairing gap.
We also notice in passing that recent microscopic calculations \cite{sces08}
based on QPs coupled to diffusive/propagating CO collective modes
demonstrate that despite the strong momentum dependence of CO
fluctuations, they can also give rise to a marginal-FL scattering in
integrated quantities (like in LDOS).

In summary, we found a wealth of evidences that dynamical order fluctuations
can be present to provide a boson-like retarded scattering mechanism.
Of course the spectral distribution of these excitations has a strong
influence on the resulting effective interaction. The last example 
of STM spectra shows that this distribution should be rather flat and featureless
(i.e. $\gamma \gg \omega_0$) so that some marginal FL self-energy 
should result. On the other hand the  presence of a marked kink feature in
ARPES spectra requires that the bosonic energy scale, at least in this
momentum and energy range,  is rather well defined.
To emphasize more clearly the role of dynamics
in the ordering fluctuations, in the rest of this paper we will rather focus on the 
effects of narrow boson spectral distributions.

\section{General formalism and self-energies}
\label{sec:formal}
In translationally invariant systems the Greens function
in real space $G_{ij}(\omega)$, corresponding to an annihilation
of a quasiparticle at site $R_i$ and subsequent creation at site
$R_j$, only depends on the difference between these
two sites, i.e. $G_{ij}(\omega)=G(R_i - R_j,\omega)$.
The same holds for the self-energy $\Sigma_{ij}(\omega)=
\Sigma(R_i - R_j,\omega)$ so that the Dyson equation
in real and momentum space reads as
\begin{eqnarray}
G_{ij}(\omega)=G_{ij}^0(\omega) +\sum_{n,m}G_{in}^0(\omega)\Sigma_{nm}(\omega)
G_{mj}(\omega) \label{eq:gf}\\
G_{\kvec}(\omega)=G_{\kvec}^0(\omega) +G_{\kvec}^0(\omega)\Sigma_{\kvec}(\omega)
G_{\kvec}(\omega) .
\end{eqnarray}
This (standard) route is followed in Sec. \ref{sec:kampf}
where we consider a self-energy $\Sigma_{\kvec}(\omega)$, derived
from the coupling of quasiparticles to a set of singular bosons.
This allows us to study the effect of a dynamical protection
of the Fermi surface in a homogeneous system close to an ordering
instability. We will show that 
the quasiparticles close to $E_F$ are not affected by the proximity to
the instability whereas at high energies the system looks ordered.

On the other hand, a description of an electronically inhomogeneous 
state, as observed in STM above some energy scale, necessarily
requires the generalization of Eq. (\ref{eq:gf}) to the case
where both, $G_{ij}(\omega)$ and $\Sigma_{ij}(\omega)$,
separately depend on sites $R_i$ and $R_j$. For simplicity we will
consider periodic modulations of the spin and charge density. We
assume that the system consists of $N_c$ non-equivalent sites which
form a supercell. These supercells repeat periodically thus generating
a Bravais lattice. The latter has a reciprocal lattice with exactly  $N_c$
{\it non-equivalent} reciprocal lattice vectors  ${\Qvec}_n$. 
For example, if the electronic order is characterized by some density
modulation with periodicity $\lambda= N_c a$
in the $x$-direction, then the reciprocal 
lattice vectors are ${\Qvec}_n=n \frac{2\pi}{\lambda} (1,0)$ with $n=1 \dots N_c$.
In this case the Dyson equation can be written in momentum space as
\begin{equation}\label{eq:elia}
{G}_{\kvec+\Qvec_m,\kvec+\Qvec_n}
=G^0_{\kvec+\Qvec_m}\delta_{m,n}
+G^0_{\kvec+\Qvec_m} 
\sum_s \Sigma_{\kvec+\Qvec_m,\kvec+\Qvec_s} 
G_{\kvec+\Qvec_s,\kvec+\Qvec_m} .
\end{equation}
The self-energy now becomes a $N_c\times N_c$ matrix where
only the diagonal elements $\Sigma_{\kvec+\Qvec_m,\kvec+\Qvec_m}$
(in case of retarded GF's) need
to obey the condition $Im \Sigma \ge 0$. While the off-diagonal
elements still obey Kramers-Kronig relation, the corresponding
imaginary part can have sign changes as a function of $\omega$.
This may lead to different low frequency behavior for the
off-diagonal $Re \Sigma$ as for the 'usual' diagonal part.

\section{Kampf-Schrieffer approach: systems without long range charge ordering}
\label{sec:kampf}
Here we follow an approach introduced by Kampf and Schrieffer
in connection with pseudogap physics due to strong AF 
fluctuations \cite{KAMPF}. 
The quasiparticles are coupled to boson excitations which are
strongly peaked in frequency and momentum space.
In our following calculations the uncoupled ground state is
a d-wave superconductor and
for simplicity the boson correlator is factorized 
in a frequency and $\qvec$-dependent part.

The self-energy is then obtained from 
\begin{equation}\label{SELF}                                               
\underline{\underline{\Sigma}}_{\kvec}(i\omega)
=-\frac{1}{\beta}         
\sum_{\qvec,ip}J(\qvec) D(ip)                                     
\underline{\underline{\tau_z G}}_{\kvec-\qvec}(i\omega-ip)             
\underline{\underline{\tau_z}}                                             
\end{equation}
where we have used Nambu-Gorkov notation so that the
unperturbed Green's function is represented by
\begin{eqnarray}
G^0_{11}(k,i\omega)&=& \frac{u_k^2}{i\omega-E_k}
+\frac{v_k^2}{i\omega+E_k}\\
G^0_{22}(k,i\omega)&=&\frac{v_k^2}{i\omega-E_k}+\frac{u_k^2}{i\omega+E_k}\\
G^0_{12}(k,i\omega)&=&G^0_{21}(k,i\omega)=-u_k v_k\left\lbrack
\frac{1}{i\omega-E_k}-\frac{1}{i\omega+E_k}\right\rbrack.
\end{eqnarray}
The BCS coherence factors are defined as $u_k^2=\frac{1}{2}
(1+\frac{\epsilon_k-\mu}{E_k})$ and $v_k^2=\frac{1}{2}
(1-\frac{\epsilon_k-\mu}{E_k})$ respectively,
and the propagator
\begin{equation}\label{dtot} 
D^{tot}(i\omega)=-\int d\nu W(\nu)\frac{ 2\nu}{(\omega^2+\nu^2)}
\end{equation}
describes the distribution of dispersionless propagating bosons.

The momentum dependent coupling is contained in the function
\begin{equation}\label{jq}
J({\bf q})=g^2 \frac{{\cal N}}{4}
\sum\limits_{\pm q_x^c;\pm q_y^c}\frac{\Gamma}{\Gamma^2
+2-\cos(q_x-q_x^c)-\cos(q_y-q_y^c)}
\end{equation}
which is enhanced at the four
equivalent critical wave vectors $(\pm q^c_x,\pm q^c_y)$.
${\cal N}$ is a suitable normalization factor introduced to
keep the total scattering strength constant while varying $\Gamma$.

We restrict
to the leading order one-loop contribution 
of the self-energy Eq. (\ref{SELF}), i.e. we replace the full by
the non-interacting Green's function on the r.h.s.
\begin{figure}[thb]
\includegraphics[width=8cm,clip=true]{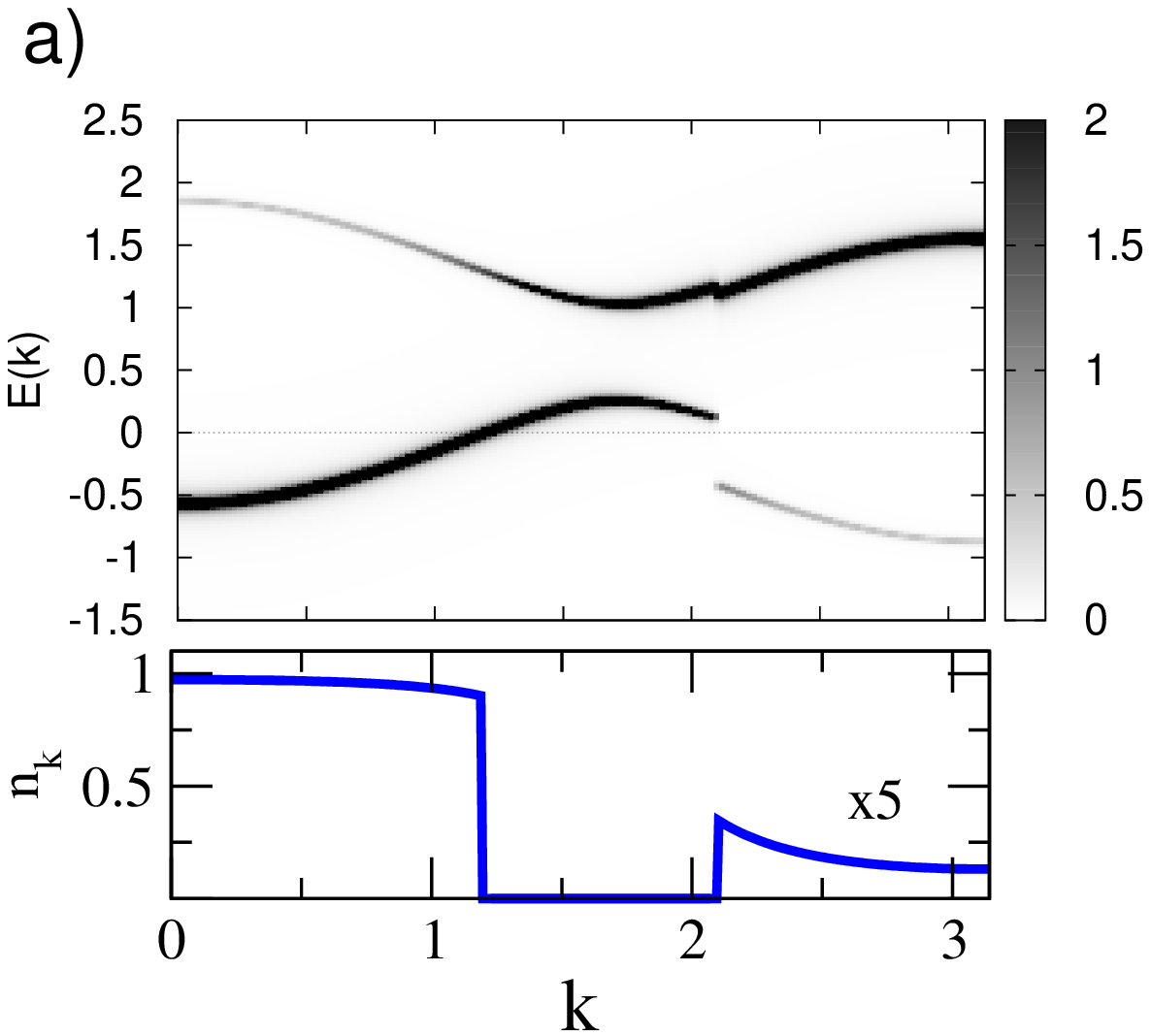}
\includegraphics[width=9cm,clip=true]{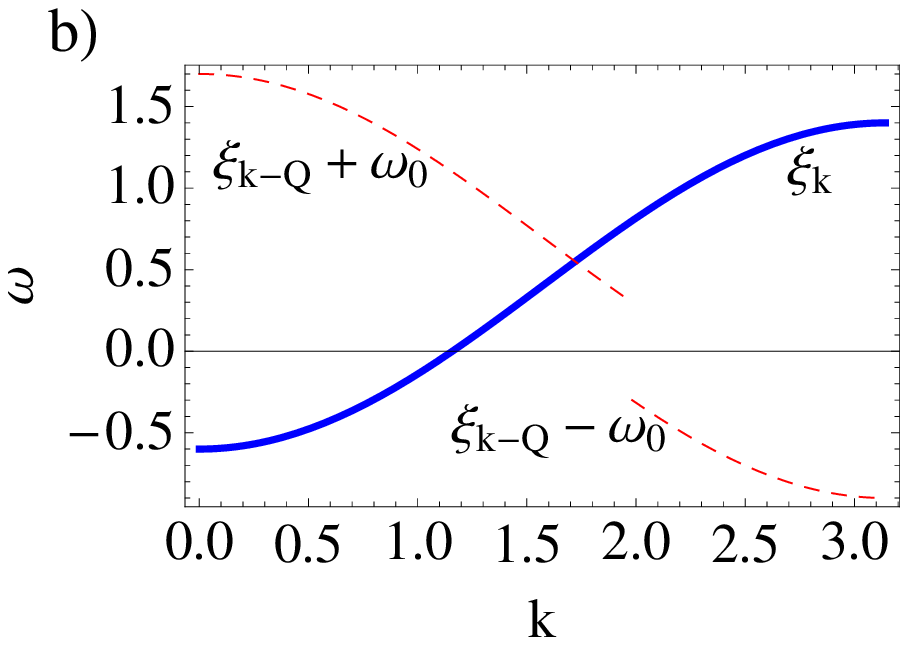}
\caption{(a)Spectral function for $\Qvec=\pi$ scattering 
in a one-dimensional model with one single boson mode
at $\omega_0=0.3$. The filling is $n=0.75$, while the 
coupling $g^2=0.15$. 
The width of the curve is proportional to the weight of the
state, and energies are measured with respect to $\mu$. 
Lower panel in (a):momentum distribution curve. 
For $k>2$ the occupation number $n_\kvec$
has been scaled by a factor of 5 to enhance the visibility.
(b) Locus of the one particle addition and removal excitation energies 
with the same parameters as in (a) but with $g=0$.}
\label{fig-1}
\end{figure}
In order to illustrate the basic features of the present approach, we
show in Figs. \ref{fig-1}-\ref{fignew5} the spectral function for 
a one-dimensional system
of electrons (dispersion $\varepsilon_k=-\cos(k)$) exposed to
dynamical CO scattering with $\Qvec=\pi$.
We start by considering the case of a single bosonic mode
oscillating at a fixed frequency $\omega_0$. Formally this corresponds to 
$W(\nu)=\delta(\omega_0-\nu)$, which was extensively discussed in 
Ref. \cite{grilli09}. 
In Fig. \ref{fig-1} (a) we report the case 
of such a single dynamical mode with a small but finite coupling 
to a system of one-dimensional electrons. To clarify the 
effect of dynamics, in panel (b) we also consider the case of a vanishingly
small coupling $g=0$.  The spectral function carries information on the
excitations of the system with one added or removed particle. For
$g=0$ the lowest energy excitations consist of an added or removed
free fermion which produce the dispersion relation indicated by the
full line in Fig.~\ref{fig-1}(b). However there are also excitations
in which the fermion is added/removed with the addition of a boson
which carries momentum $\Qvec=\pi$ and energy $\omega_0$. The
dispersion relation of these excitations is depicted by the dashed
line. For $g=0$ all the weight is in the main band labeled $\xi_k$. 
The effect of a
finite coupling $g$ is to give some spectral weight to the shadow band
at $\xi_{k-Q}\pm\omega_0$ and to introduce some level repulsion when the
bands cross. The important point is that in contrast to a really
statically ordered system the shadow band never touches the Fermi level,
but it is separated from it by the energy to create the bosonic
excitation. 
We associate the main band with the QP band and 
the shadow band with the 
incoherent spectral weight arising from the scattering with CO fluctuations. 
Clearly close enough to the Fermi level only the QP band exists but
the high energy spectral function resembles that of an ordered
system. 
Furthermore, the momentum distribution function is practically
identical to that of an ordered system, so that a naive analysis of a direct 
measurement of $n_{\bf k}$ by Compton scattering or by integrating the
photoemission spectral function will lead to a different conclusion on 
the ordering of the system than a low energy photoemission experiment.

The above example illustrate in a simple way the main idea of a
protected Fermi surface in an almost ordered system. On the other hand,
the assumption  $\Gamma \to 0$  is too singular from the point of view
of Fermi liquid theory and yields unphysical results if the
interaction is increased beyond a certain value or in the case of nesting. 
 For larger coupling
the backbending of the main band in Fig. \ref{fig-1} will induce
an additional FS crossing and thus the generation of spurious 'shadow features'
at the Fermi level.
In Ref. \cite{grilli09} we have circumvented this complication by
the introduction of phenomenological vertex corrections aimed
to suppress the coupling to the boson at the Fermi energy.
Here we show that the problem can also be avoided by employing
a more physical coupling which is less singular.  
Fig. \ref{fig1d}a reports the spectral function for the same parameters
as Fig. \ref{fig-1} but with a finite correlation length 
(i.e. finite $\Gamma=0.1$ in Eq. (\ref{jq})).
Clearly the backbending
of the main dispersion is significantly reduced thus derogating the
tendency towards shadow feature formation at the Fermi level. On the
other hand the high energy states still resemble the structure of a $\Qvec=\pi$
ordered state with of course somewhat smaller weight as
compared to Fig. \ref{fig-1}. Furthermore the momentum distribution
function shown in the lower panel, still resembles that of an ordered
system, with a momentum broadening of the shadow features similar to the
momentum broadening of the small Fermi surface of the heavy-fermion
system of Fig.~\ref{fig-HF}. 

Within our scenario a finite correlation length is especially
important in the case of nesting. Here the singular scattering
with $\Gamma \to 0$ would produce a gap in the spectrum
for those momenta which fulfill the nesting condition as shown
in Fig. \ref{fignew5}a for the half-filled one-dimensional case.
This of course would cause severe problems in two dimensions
where parts of the Fermi surface can always be nested by an 
appropriate scattering vector.
Fig. \ref{fignew5}b reveals that also in this case a finite correlation
length introduces a quasiparticle band crossing the Fermi energy
while the high energy part of the spectrum still resembles the structure
of an ordered state.

We now move to the case of a distribution
of bosons which can be tuned to be more or less broad.
Our results on cuprates in the second part of this section
are obtained for a linear distribution of bosonic modes up to
some cutoff energy $\omega_{max}$, i.e. $W(\nu)$ in Eq. (\ref{dtot})
is given by $W(\nu)=\left(2\nu/\omega^2_{max} \right)\Theta(\omega_{max}-\nu)$.
Note however, that the basic features of the spectra do not depend 
significantly on this choice, but similar results would be obtained 
by a lorentzian distribution of bosonic modes.

In order to exemplify the influence of a frequency  broadening on the spectra, 
Fig. \ref{fig1d}b reports the
case of $\omega_{max}=0.3$ for a doped one-dimensional
system which can be compared with the single frequency case reported
in Fig. \ref{fig-1}.
\begin{figure}[thb]
\includegraphics[width=8cm,clip=true]{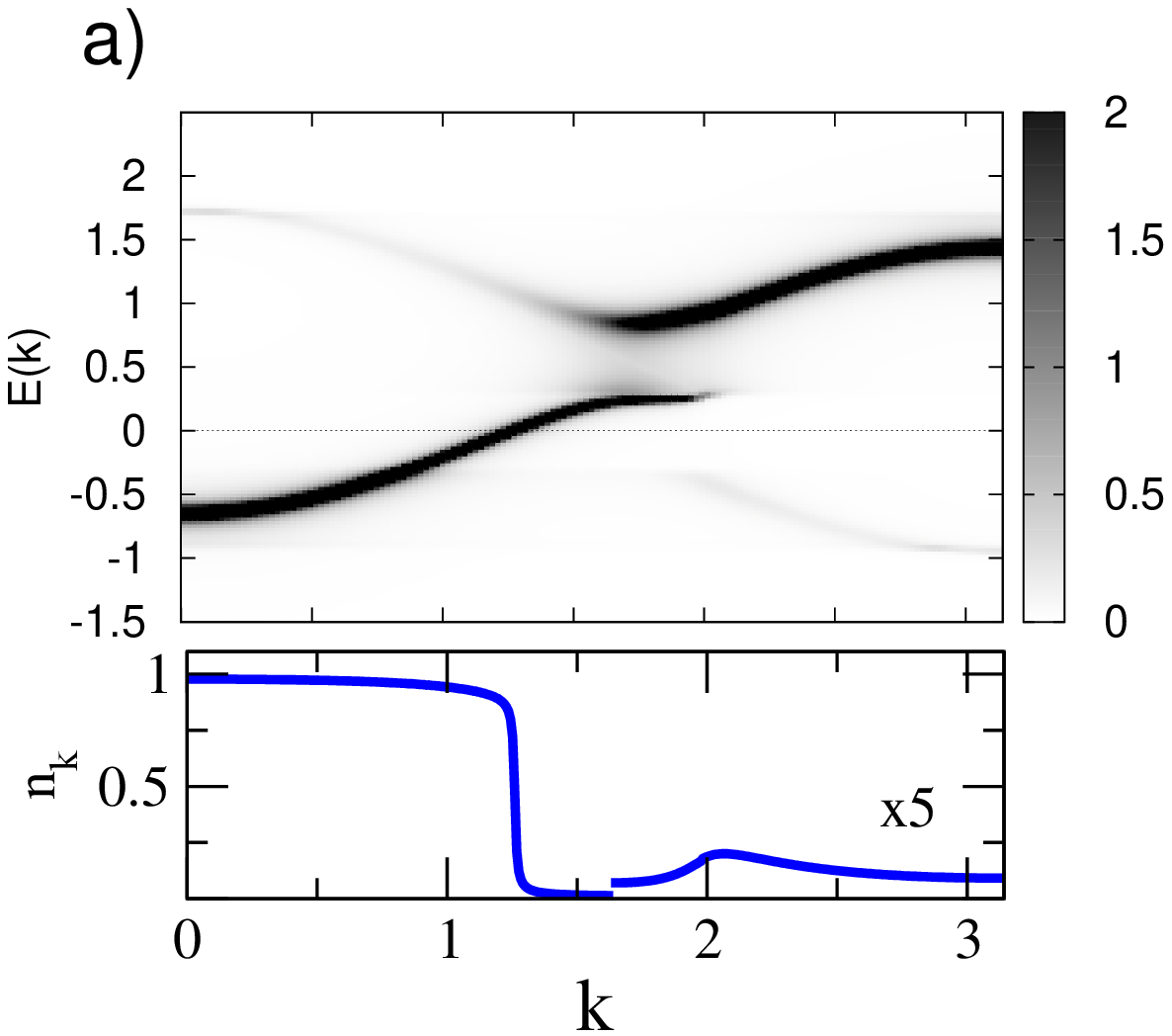}
\includegraphics[width=8cm,clip=true]{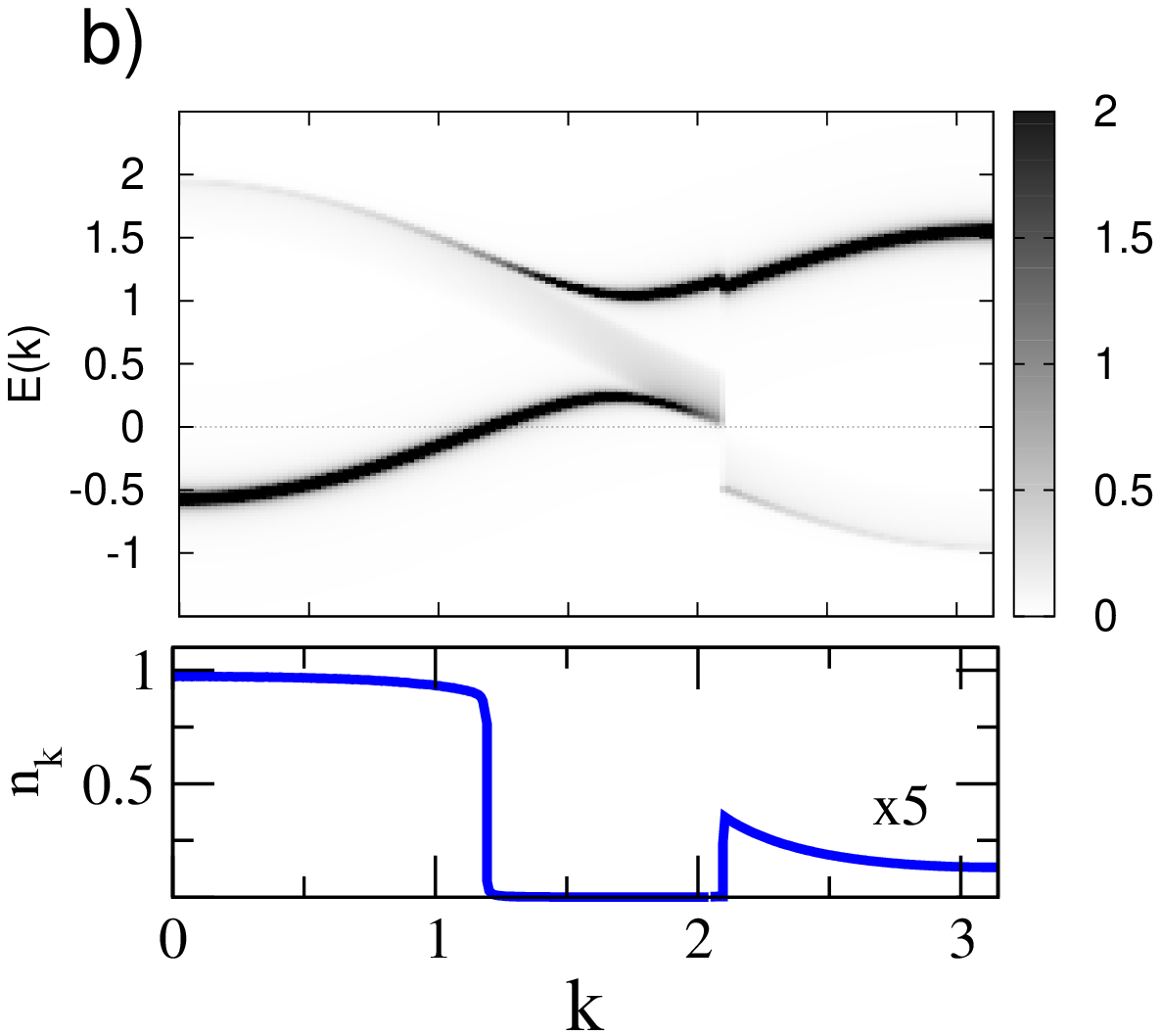}
\caption{Spectral function for $\Qvec=\pi$ scattering 
in a one-dimensional model. The electron density is $n=0.8$ and 
coupling $g^2=0.1$. 
(a) Finite correlation length $\Gamma=0.1$ but single boson frequency 
$\omega_0=0.3$.
(b) Linear distribution of boson frequencies with $\omega_{max}=0.3$
but infinite correlation length $\Gamma \to 0$.
Lower panels in: momentum distribution curve. 
For $k>\pi/2$ the occupation number $n_\kvec$
has been scaled by a factor of 5 to enhance the visibility.}
\label{fig1d}
\end{figure}
Also in this case, for the doped system %[panel (a)] 
the spectra at 
higher energy resemble those of an ordered system due to the
appearance of the shadow bands, but again
these shadow bands do not cross the Fermi level
and are gapped on a scale of the bosonic excitation energy $\omega_{max}$. 
Therefore close
to the Fermi level the system is protected by the scattering
which only becomes apparent on an energy scale larger than that
 of the bosons.
As can be seen from the lower panels of Fig.~\ref{fig1d} both in the
case of momentum broadening as well as frequency broadening the
protected Fermi-liquid like  quasiparticles coexist with a momentum
distribution function resembling that of a system with long-range order.
This thus
mimicks the results obtained by  Zhou {\it et al.} in cuprates 
using  ARPES.\cite{zho99,zho01}

\begin{figure}[thb]
\includegraphics[width=8cm,clip=true]{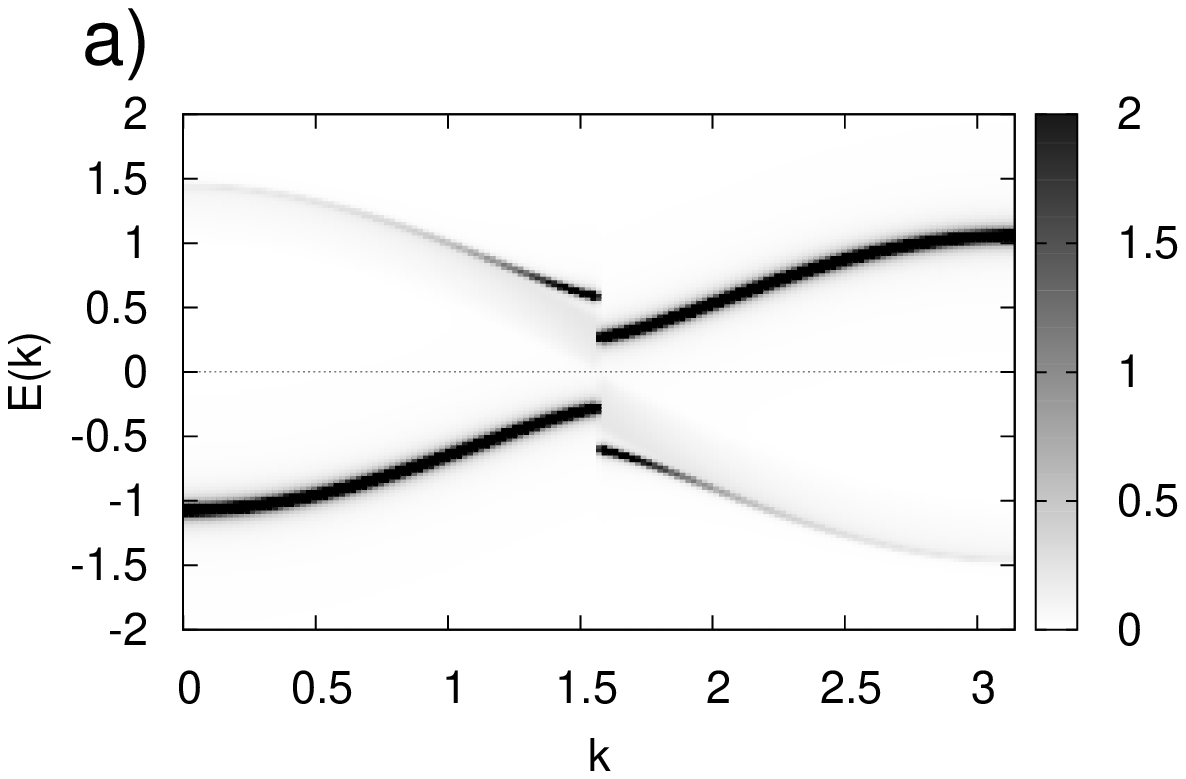}
\includegraphics[width=8cm,clip=true]{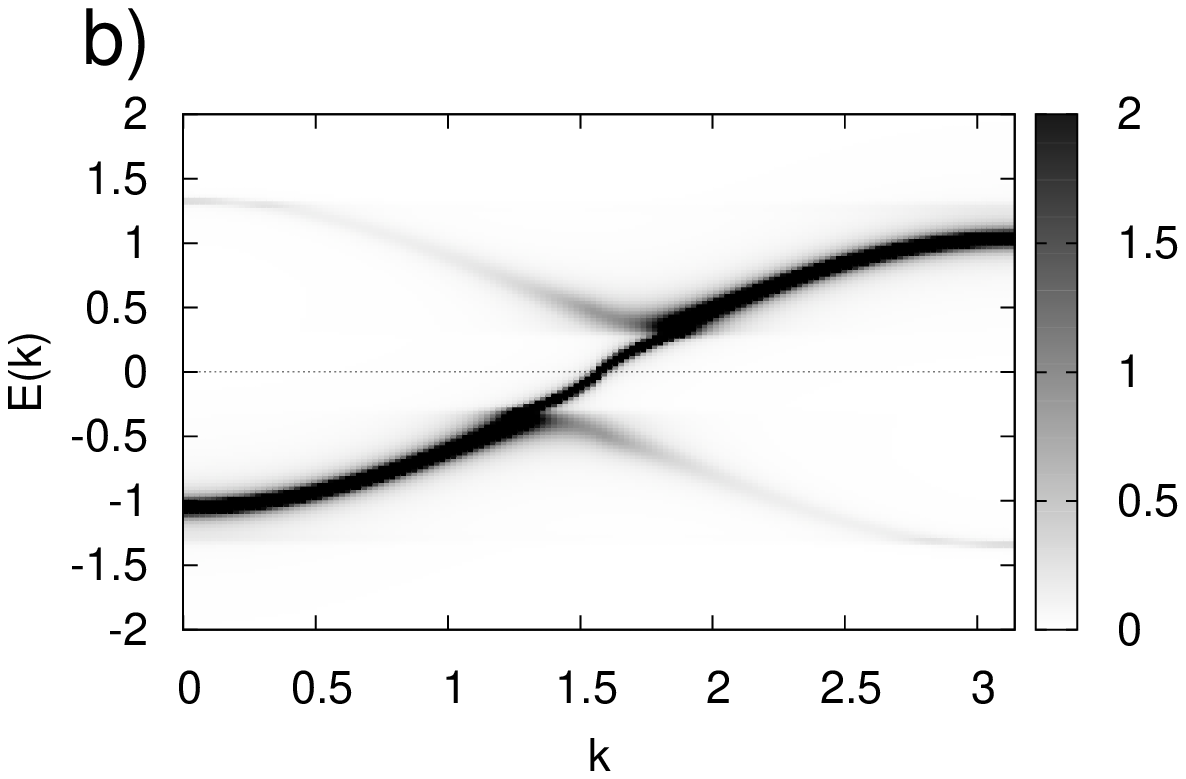}
\caption{Spectral function for $\Qvec=\pi$ scattering 
in a half-filled one-dimensional model. Coupling $g^2=0.1$
and boson frequency  $\omega_0=0.3$.
(a) Infinite correlation length $\Gamma \to 0$.
(b) Finite correlation length $\Gamma=0.1$.}
\label{fignew5}
\end{figure}

We proceed by  applying the Kampf-Schrieffer-type approach to
the investigation of quasiparticle
spectra in the superconducting state of high-T$_c$ superconductors.
In this context we ask the question how this scattering
affects the effective spectral gap of the system and in which
way it influences the quasiparticle weight.
Our investigations are based on a parametrization of the
dispersion of Bi2201 from Ref. \cite{marki10} 
\begin{eqnarray}
\varepsilon_k&=&-2t[\cos(k_x)+\cos(k_y)]-4t'\cos(k_x)\cos(k_y)\nonumber \\ 
&-&2t''[\cos(2k_x)+\cos(2k_y)]-u \label{eq:disp}
\end{eqnarray}
with $t=217.5 meV$, $t'=-60 meV$ and $t''=20 meV$. 
The chemical potential is $u=-0.23 eV$ corresponding to a
particle number of $n\approx 0.81$.
The CO scattering vectors are restricted to 
 $\Qvec_m=(\pm 0.85 1/a,0),(0,\pm 0.85 1/a)$
which connect the FS segments at the
antinodal $(0,\pi)$  and  $(\pi,0)$ points, respectively. 
For simplicity, the results below are derived for the case of infinite 
correlation length (i.e. $\Gamma \to 0$).
However, the model can be extended to CO scattering
with finite correlation length which from a technical point of view
requires some more numerical effort.  
To circumvent the $\Gamma\to 0$ complications associated with 
nesting in the two-dimensional case, we will 
concentrate
on the superconducting ($d$-wave) state so that the FS is gapped  
in those regions of momentum space where the scattering could fulfill
the  nesting condition. Please remember that these complications
are avoided by a finite $\Gamma$.
The SC gap is approximated by a simple harmonic $d$-wave structure
$\Delta(\kvec)=\Delta_0[\cos(k_x)-\cos(k_y)]$ with $\Delta_0=30 meV$
and the linear distribution of boson modes is cut off at $\omega_{max}=0.1 eV$.
\begin{figure}[htb]
\includegraphics[width=8cm,clip=true]{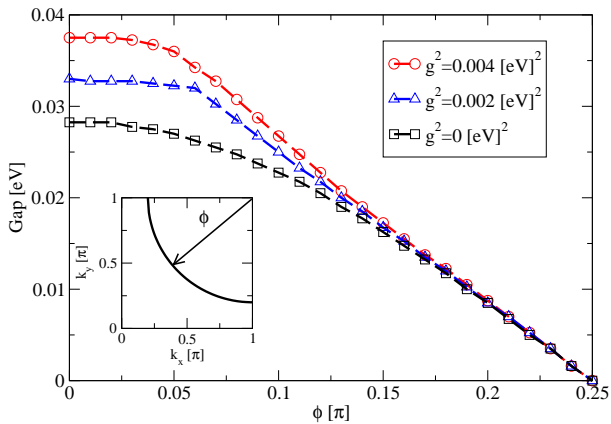}
\caption{Minimum gap deduced from the 'minimum gap' locus
for various couplings of the dynamical CO scattering.}                         \label{figgap}
\end{figure}

The evolution of the SC gap along the underlying FS 
is shown by the square symbols in Fig. \ref{figgap} 
(plotted as a function of the angle $\phi$ defined in the inset).
It follows approximately the relation $\Delta(\phi)=\tilde\Delta_0
\cos(2\phi)$ with $\tilde\Delta_0=28.5 meV < \Delta_0$ due
to the displacement of the Fermi segments from the M-points.
Switching on the CO scattering alters the underlying FS
and we have to find the effective minimum gap by scanning over the
whole Brillouin zone. Fig. \ref{figgap} shows that the scattering
induces a deviation from the simple harmonic $d$-wave structure 
and increases the effective gap upon approaching the M-point.
We can fit the resulting angular gap dependence by including
higher harmonics which for the coupling $g^2=0.004 (eV)^2$
(circles in Fig. \ref{figgap}) leads to
$\Delta(\phi)=\tilde\Delta_0 [0.9\cos(2\phi) + 0.1\cos(6\phi)]$ 
with $\tilde\Delta_0=38 meV$.
This behavior is close to the observed midpoint shift of the EDC
leading edge in recent ARPES experiments \cite{kondo07} where
an even larger anharmonicity was reported. Whereas we could easily
reproduce larger effective gaps around the M-points this would
put in jeopardy our weak coupling
approach. We, however, note that the deviation from the harmonic
gap structure in our calculations occurs at essentially the same
angle ($\approx 0.15 \pi$) as in the experiment reported
in Ref. \cite{kondo07}.
The anharmonicity of the gap function is also supported by
STM experiments on underdoped Bi$_2$Sr$_2$CaCu$_2$O$_{8+\delta}$ \cite{pushp09} 
where the fitting of the local density of states (LDOS) requires 
an angular dependence of the superconducting gap which is composed
of a harmonic $d$-wave and an additional
anharmonic contribution which contributes in the antinodal regions.

\begin{figure}[thb]
\includegraphics[width=9cm,clip=true]{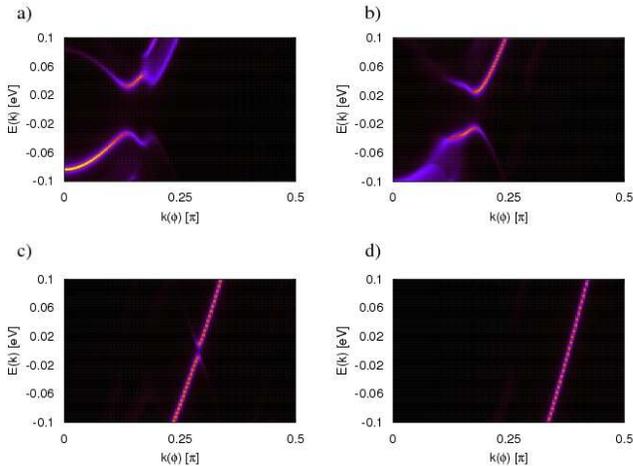}
\caption{Spectral function along selected radial cuts of the Brillouin zone.  
(cf. inset to Fig. \ref{figgap} as a function of the $k_x$ projection. 
From a) to d) $\phi/\pi=0, 0.1, 0.2, 0.25$.
The intensity scale corresponds to $\frac{1}{\pi}Im G^{11}_\kvec(\omega)\delta
\omega$  
Parameters: $g^2=0.002 (eV)^2$, $\Delta_0=30 meV$.}              
\label{figakw}
\end{figure}

Fig. \ref{figakw} reports the spectral function for selected
points along the 'minimum gap' locus. Since the scattering momentum
connects the antinodal segments of the underlying FS the
corresponding spectrum (panel a) displays the weak incoherent shadow features 
both below and above $E_F$, which are additionally shifted in  energy
by the mixing to the boson excitations. Upon moving away from the
antinodes (panel b) the scattering predominantly affects states below
E$_F$ so that the associated high energy spectrum acquires large
incoherent characteristics.
On the other hand the nodal regions (panels c,d) are hardly affected
by the scattering and the corresponding dispersion resembles that
of a 'clean' $d$-wave superconductor.

We proceed by calculating the spectral
weight of the SC coherence peaks in different parts of the
Brillouin zone. The corresponding experimental ARPES study
has revealed a strong angular dependence of the weight
upon scanning from the antinodes towards the nodal region \cite{kondo09}.
In this work the (symmetrized) energy distribution curves for 
$\kvec=\kvec_F$ ($\equiv$ Fermi momentum) at temperatures
slightly above T$_c$, $\rho^>_{\kvec_F}(\omega)$, have been 
substracted from those obtained below T$_c$, $\rho^<_{\kvec_F}(\omega)$.
The coherent spectral weight $W_{CP}$ is then defined as the
integral over the positive area of this difference, i.e.
\begin{equation}
W_{CP}=\int_{-\infty}^{\omega_c} d\omega\left(\rho^<_{\kvec_F}(\omega)-
\rho^>_{\kvec_F}(\omega) \right)
\end{equation}
and $\omega_c$ is defined as the binding energy
where the integrand becomes negative close to $E_F$.
Analysis of the ARPES data \cite{kondo09} revealed a characteristic
angular dependence of $W_{CP}$ (Fig. 2h in Ref. \cite{kondo09}) 
which vanishes at the node and is
additionally suppressed around the antinodes. The suppression
increases with underdoping so that $W_{CP}$ develops a (doping-dependent)
maximum at intermediate angles between nodal and antinodal regions.

In order to understand these results let us first start with the
case of an 'unperturbed' ideal superconductor.
Here the difference between SC and normal state
spectral function on the Fermi surface    
$
\delta Im G_{{\bf k}}(\omega)= Im 
G^{SC}_{\bf k}(\omega)- Im G^{NS}_{\bf k}(\omega)$, consists of two
delta-peaks at $\omega=\pm \Delta({\bf k})$ 
with weight one-half and a negative delta-peak at $\omega=0$ with weight
one, respectively. Thus the coherent spectral weight $W_{CP}$ as the integral
over the ($\omega \le 0$) coherence peak yields   $W_{CP}=1/2$ for
a finite SC gap and $W_{CP}=0$ for $\Delta=0$.
In case of disorder the delta-peaks are broadened into lorentzians
(width $\delta$) for which an analogous consideration yields 
\begin{eqnarray}
W_{CP}({\bf k})&=&\frac{1}{\pi}\arctan\left(\sqrt{1+\alpha_{\bf k}^2}\right)
- \frac{1}{2\pi}\arctan\left(\sqrt{1+\alpha_{\bf k}^2}+2\alpha_{\bf k}\right)\nonumber \\
&-& \frac{1}{2\pi}\arctan\left(\sqrt{1+\alpha_{\bf k}^2}-2\alpha_{\bf k}\right)\label{atan}
\end{eqnarray}
with $\alpha_{\bf k}=\Delta_{\bf k}/\delta$. 
The function Eq. (\ref{atan}) is shown in
the inset of Fig. \ref{figwqp} and suggests that the slope of the
observed angular dependence of $W_{CP}$ around the nodes is probably
determined by the amount of disorder in the sample.
On the other hand we can attribute the suppression of $W_{CP}$
around the antinodes to the influence of the fluctuating CO order
which predominantly affects these regions in momentum space.
\begin{figure}[htb]
\includegraphics[width=7cm,clip=true]{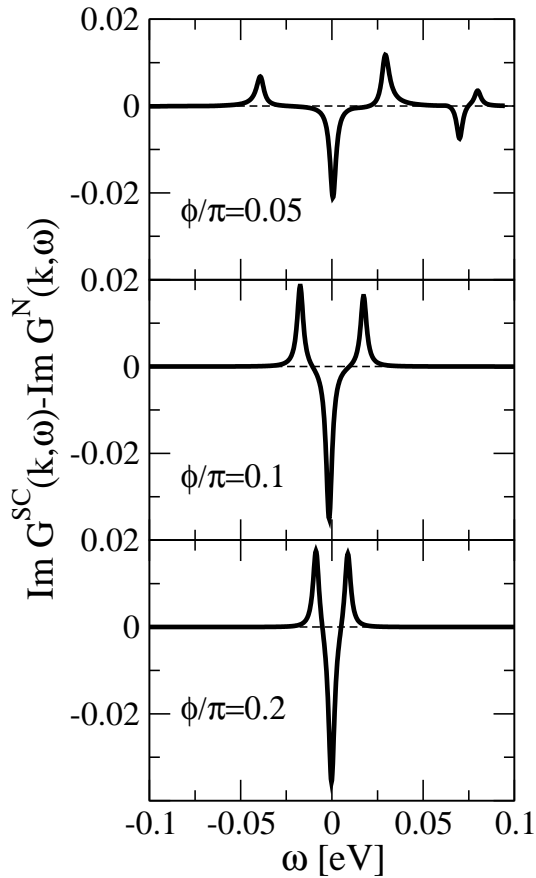}
\caption{Difference between EDC's in the SC and normal state 
for selected momenta along the
'minimum gap' locus. From top to bottom $\phi/\pi=0.05, 0.1, 0.2$.
Parameters: $g^2=0.004(eV)^2$, $\Delta_0=30 meV$.}
\label{figedc}
\end{figure}
Fig. \ref{figedc} shows the difference between normal state and SC EDC curves
(both under the presence of dynamical CO scattering) for selected momentum
points along the minimum gap locus. The suppression of coherent spectral weight
around the antinodes due to the scattering is clearly apparent.
On the other hand, close to the nodes the coherence peaks shift to small
energies and due to the broadening the associated weight is partially
absorbed by the negatively weighted normal state peak causing the
vanishing of $W_{CP}$ for $\phi \to \pi/4$.   

We are now in the position to theoretically model the angular dependence
of $W_{CP}$ as shown in Fig. 2h of the ARPES study of Ref. \cite{kondo09}. 
We select parameters in such a way that the overdoped sample is
described by an unperturbed $d$-wave superconductor, but with significant
broadening of the spectral function due to the dopant induced disorder.
Optimally doped and underdoped samples are modelled by a smaller
amount of disorder but with increasing coupling to the CO fluctuations.
The resulting angular dependencies of $W_{CP}$ are shown in Fig. \ref{figwqp}
and qualitatively reproduce the experimental ones of Ref. \cite{kondo09}.
It should be noted that the experimental data of Ref. \cite{kondo09} report
a complete suppression $W_{CP}$ around the antinodes for the underdoped sample.
Within the present approach this would require much larger couplings
beyond our weak coupling computation.
However, it should also be mentioned that the suppression may not be as
strong as the analysis of Ref. \cite{kondo09} suggests. In fact, as discussed
above, their method of extracting $W_{CP}$ is based on the {\it positive} area
difference between EDC's below and above T$_c$ which (since spectral weight is 
conserved) should be the same as the {\it negative} area difference which
has not been analyzed. Since the latter is still 
substantial in both optimally doped and underdoped samples around the 
antinodes (cf. Figs. 2d,f in Ref. \cite{kondo09}) this
may point towards a less complete suppression than what is suggested by
Fig. 2h of Ref.\cite{kondo09}.

\begin{figure}[thb]
\includegraphics[width=9cm,clip=true]{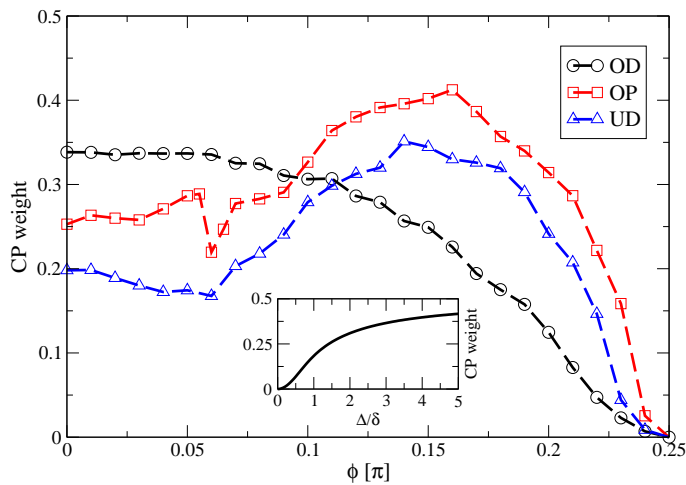}
\caption{Model for the weight of the coherence peak along
the 'minimum gap' locus. Overdoped (OD):
$g^2=0 [eV]^2 $, $\Delta_0=20 meV$, $\delta=0.04 eV$.
Optimally doped (OP): 
$g^2=0.003 [eV]^2 $, $\Delta_0=30 meV$, $\delta=0.02 eV$.
Underdoped (UD):
$g^2=0.005 [eV]^2 $, $\Delta_0=20 meV$, $\delta=0.02 eV$.
The inset shows the function derived in Eq. (\ref{atan}.) }
\label{figwqp}
\end{figure}
Note that the dip feature around $\phi \approx 0.05$ in the curve 
for optimal doping (square symbols) results from the scattering 
of near-antinodal points between adjacent Brillouin zones.
As can be seen this dip is smeared out for larger coupling and
is obviously a result of our single momentum CO scattering
approximation.
We expect that the implementation of finite range correlations
by e.g. a lorentzian distribution of scattering vectors would
wash out the features associated with nesting, however, the 
mean magnitude should still be comparable with the
separation of the antinodal FS segments so that the scattering still 
induces the suppression of $W_{CP}$ in that
area of the Brillouin zone.

\section{Phenomenological approach with long range charge ordering}
\label{sec:phen}
While the Kampf-Schrieffer limit captures in principle the
dichotomy between low energy protected quasiparticles and
high energy CO scattering, it preserves the homogeneity of
the ground state. 
However, STM experiments show that the real space
spectra taken at different sites are not equivalent often showing a glassy
character. Even more, as outlined in the introduction,
the observation of almost dispersionless LDOS modulations at
higher energy \cite{kohsaka08}
provides additional evidence for real translational symmetry breaking. 
In this case a translational invariant treatment is
clearly inappropriate.  

For simplicity we will restrict to system in which the symmetry
breaking is periodic but the same physics applies to disordered systems.
As discussed in Sec. \ref{sec:formal},  symmetry breaking 
 is described by the off-diagonal GF and self-energy
contributions in Eq. (\ref{eq:elia}) which will be phenomenologically
constructed in the present section.

For this purpose we consider particles on a lattice which gain kinetic energy
from hopping processes ($\sim t_{ij}$) but are additionally scattered
by an inhomogeneous, local, and frequency dependent self-energy. 
Thus in Eq.~\ref{eq:elia} we set 
$\Sigma_{ij}(\omega) \to \Sigma_{i}(\omega)\delta_{ij}$ and Dyson
equation in real space reads, 
\begin{equation}\label{eq:real}
\left(\omega - \Sigma_i(\omega)\right)G_{ij}^\sigma = \delta_{ij}              
+  \sum_{p}t_{ip}G_{pj}^\sigma .                                                
\end{equation}                                                             

\begin{figure}[thb]
\begin{center}
\includegraphics[width=8cm,clip=true]{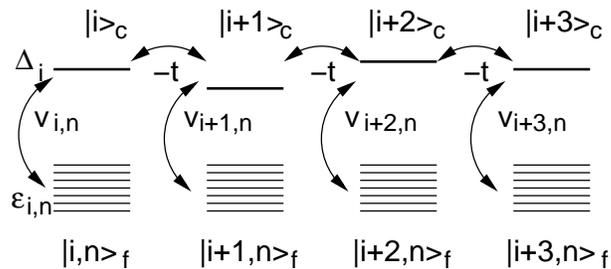}
\end{center}
\caption{Schematic structure a 1D model hamiltonian
 implementing dynamical CO order. The itinerant 'c'-states represent the
non-interacting quasiparticles of the system which are coupled  
to a bath of f-states representing collective charge or spin fluctuations.
Static CO corresponds to local shift of the 'c'-state onsite
levels by $\Delta_i$.}
\label{fig1}
\end{figure}

Our objective in the following is to construct a phenomenological
self-energy $\Sigma_i(\omega)$ which leaves the states around $E_F$
protected from the scattering but is consistent with some kind
of electronic order at higher binding energies.
Note that the local charge density at site $m$ is determined by
the sum over all frequencies of the off-diagonal GF 
\begin{equation}
\langle n_m\rangle = \frac{1}{\beta N}\sum_{\kvec,s,ip}\exp(i\Qvec_s \Rvec_m)
G_{\kvec,\kvec+\Qvec_s}(ip)
\end{equation}
where the $\Qvec_s$ are reciprocal lattice vectors introduced in Sec. \ref{sec:formal}. 
One is thus able to construct a $\Sigma_i(\omega)$
which vanishes for $\omega \to 0$ (and thus also the off-diagonal GF)
but causes a inhomogeneity in $\langle n_m\rangle$ via its high
frequency part.

In order to accomplish this task it is necessary to maintain the
correct analytical properties of the self-energy and GF. Without loss
of generality we can introduce a pole expansion for the self-energy,  
\begin{equation}\label{eq:self}
\Sigma_i(\omega)=v_i^2\sum_n\frac{1}{\omega-\epsilon^f_{n,r}}+\Delta_i
\equiv v_i^2 f_i(\omega)+\Delta_i. 
\end{equation}
with the constant $\Delta_i$ controlling the high frequency limit.
The expansion in Eq.~(\ref{eq:self}) coincides with the self energy of
the following Fano-Anderson Hamiltonian,
\begin{equation}\label{eq:aux}
H^{aux}=H^0 + \sum_{i,n,\sigma}v_{i,n} \left\lbrack c_{i\sigma}^\dagger f_{in,\sigma}
+ h.c. \right\rbrack + \sum_{i,n,\sigma}\varepsilon_{in}f^\dagger_{in,\sigma}
f_{in,\sigma}
\end{equation}
and which is illustrated in Fig. \ref{fig1} for a one-dimensional system.
Besides the hopping between nearest neighbor sites ($\sim t$), itinerant
(c) electrons (representing the quasiparticles) on
each site can be transfered ($\sim v_{i,n}$) to a bath of auxiliary f-states
('impurities') representing the collective fluctuating charge or 
spin environment which scatters the itinerant carriers. 
It is straightforward to show that the GF for the ($c$) quasiparticles
obeys Eq. (\ref{eq:real})
and the local self-energy is determined by the coupling to the               
$f$-level distribution. The mapping to
this auxiliary hamiltonian problem ensures that the phenomenological
self-energy has all the correct analytical properties.

The function $f_i(\omega)$
describes the dynamics of the CO and depends on the site and frequency.
 $\Delta_i$ describes the frequency independent modulation of onsite levels.

Since the self-energy is local, upon Fourier transforming 
it no longer
depends on $\kvec$ but only on the transfered momentum, i.e.
$\Sigma_{\kvec+\Qvec_m,\kvec+\Qvec_n} \to \Sigma_{\Qvec_m-\Qvec_n}$.
Thus the diagonal elements of the self-energy matrix are equal and given by
$\Sigma_{0}$ while other values of the transfered momentum correspond to 
off-diagonal elements.

To gain some insight into the properties of this approach
consider a half-filled one-dimensional model 
(dispersion $\varepsilon_k=-\cos(k)$) with period
doubling ($N_c=2$). 
In this case the Fourier transformed 
$\Sigma_Q(\omega) = 1/N \sum exp(iQ R_j)\Sigma_j(\omega)$ has only
two non-vanishing components, $\Sigma_{Q=0}(\omega)$ (diagonal) and
$\Sigma_{Q=\pi}(\omega)$ (off-diagonal). 

With regard to the auxiliary hamiltonian Eq. (\ref{eq:aux}) 
there are essentially two possibilities to
realize the dynamical modulation: An alternating
coupling to a site independent frequency spectrum (in the following
refered to as 'variant 1') or
a constant coupling to a frequency spectrum which alternates
from site to site ('variant 2').

Consider first variant 1. 
The (site-independent) frequency spectrum is assumed to be peaked
at $\omega=\pm \omega_0$ and reads as 

\begin{equation}\label{eq2p}
f(\omega)=\frac{1}{\omega-\omega_0} + \frac{1}{\omega+\omega_0} .
\end{equation}

Coupling only odd sites ($v^2_{2i}=0$, $v^2_{2i+1}=2v^2$) to the
spectrum yields the self-energies
\begin{equation}
\Sigma_{Q=0}(\omega)=-\Sigma_{Q=\pi}(\omega)=v^2f(\omega).
\end{equation}
In the weak coupling limit ($\lambda \equiv v/\omega_0 \ll 1$)
we can expand the resulting Green's function in the limits
$\omega > \omega_0$ and $\omega < \omega_0$
which yields
\begin{eqnarray}\label{eq:expand}
G_k(\omega) &=& \frac{Z}{\omega - Z\varepsilon_k} \hspace*{1.95cm}
\mbox{for} \hspace*{0.2cm} \omega < \omega_0 \\
G_k(\omega) &=& \frac{\omega+\varepsilon_k-\frac{2g^2}{\omega}}
{(\omega-E_k)(\omega+E_k)} \hspace*{0.5cm}
\mbox{for} \hspace*{0.2cm} \omega > \omega_0 \nonumber
\end{eqnarray}
with $Z=1/(1+\lambda)$ and $E_k=\sqrt{\varepsilon_k^2+4v^2}$.
\begin{figure}[htb]
\includegraphics[width=9cm,clip=true]{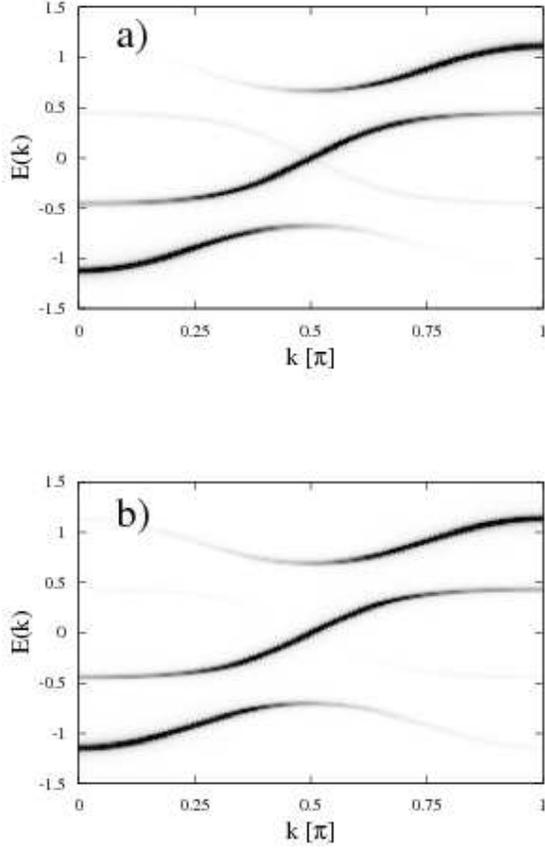}
\caption{Spectral function for a one-dimensional model
with dynamical CO scattering with period $q=\pi$ 
for a two-pole self-energy.
Top panel: alternating coupling to a constant frequency spectrum;
Lower panel: constant coupling to a site-dependent frequency spectrum.
Parameters: $v^2=0.05$, $\omega_0=0.5$}
\label{figph1d}
\end{figure}
Fig. \ref{figph1d}(a) displays the resulting spectral function for variant 1.
The spectrum is composed of the low energy part ($\omega < \omega_0$)
with renormalized dispersion and quasiparticle weight $(\sim Z$).
Also apparent is the weak shadow band due to the low energy
scattering induced by $\Sigma_{Q=\pi}=\lambda^2\omega$.
At high energy ($\omega > \omega_0$) the electronic structure 
changes to that of a CDW with poles at $\pm E_k$, i.e.
a CDW which is given by $\Delta=2 v$.

Consider now the variant 2 with a constant coupling 
to an alternating frequency spectrum.
For simplicity we assume that the latter 
is determined by a pole at $\omega=(-1)^{R_i}\omega_0$ so that
the self-energies read as
\begin{eqnarray}
\Sigma_{Q=0}&=&v^2 \left\lbrack \frac{1}{\omega-\omega_0}
+\frac{1}{\omega+\omega_0}\right\rbrack \label{eq:as2} \\
\Sigma_{Q=\pi}&=&v^2 \left\lbrack \frac{1}{\omega-\omega_0}
-\frac{1}{\omega+\omega_0}+\frac{2}{\omega_0}\right\rbrack. \label{eq:s2}
\end{eqnarray}
Note that a static component $\Delta=\lambda^2\omega_0=\frac{v^2}{\omega_0}$ has
been added to the finite momentum self-energy in order to 
keep the limit $\Sigma_{Q=\pi}(\omega=0)=0$.

The difference to the previous case is first in the low energy behavior
of $\Sigma_{Q=\pi}$ which now is of the order ${\cal O}(\omega^2)$ \cite{note}.
This further reduces the intensity of the low energy shadow bands
which is apparent from the spectral function shown in
Fig. \ref{figph1d}(b). Second, the high-energy part of the 
off-diagonal self-energy Eq. (\ref{eq:s2}) approaches 
$\Sigma_{Q=\pi}(\omega\to \infty)=2v^2/\omega_0$ and thus
corresponds to the self-energy of the static solution. Thus in contrast to
variant 1, which for the present choice of $f(\omega)$ (cf. Eq. (\ref{eq2p}))
leads to a vanishing CO scattering in the limit $\omega\to \infty$, 
variant 2 leads
to a finite off-diagonal Green's function $G_{\kvec,\kvec+\Qvec}$ 
for all energies $\omega>>\omega_0$.
It should be noted that in case of variant 2 the weak coupling expansion of 
the {\it diagonal} Green's function $G_{\kvec,\kvec}$ at low and high energies 
leads to the same results as before [cf. Eq. \ref{eq:expand})].

\begin{figure}[htb]
\includegraphics[width=8cm,clip=true]{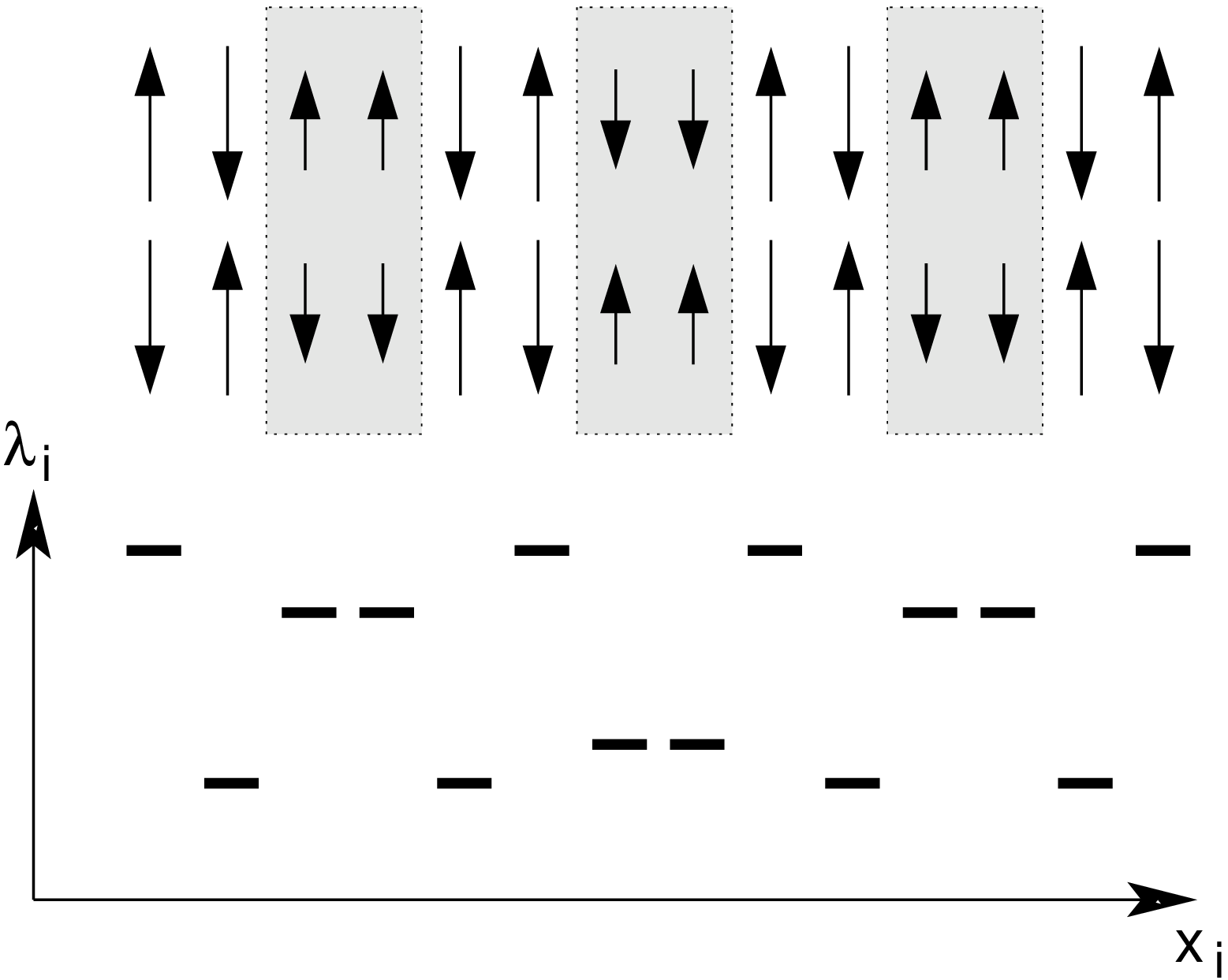}
\caption{Sketch of the bond-centered stripe structure with
hole enriched domain walls indicated by the shaded areas.
The lower part shows the variation of the local chemical
potential perpendicular to the stripe direction.
}
\label{figlevel}
\end{figure}

Being phenomenological in nature we do not have a microscopic electron-electron
or electron-boson model from which
this self-energy can be derived (contrary to the Kampf-Schrieffer
self-energy which can be derived from a fermion-boson model).
However, we can motivate the above self-energy variants from a more
microscopic point of view by considering the coupling of
a local charge at site ${\bf R}_i$ 
(operator $c_i^{(\dagger)}$, site energy $\varepsilon_i$, coupling constant 
$g_i$) 
to a bosonic excitation (operator $b_i^{(\dagger)}$, frequency $\omega_0$). 
If the site is empty, the lowest order addition states are 
obtained from $c_i^\dagger |0\rangle$ and $c_i^\dagger b_i^\dagger |0\rangle$     
with addition energies $\varepsilon_i$ and $\varepsilon_i+\omega_0$, 
respectively.
The addition self-energy thus becomes 
$\Sigma^+(\omega)=g_i^2 (1-\langle n_i\rangle )/(\omega-\varepsilon_i-\omega_0)$.
Analogously the lowest order removal states are obtained from
$c_i |i \rangle$ and $c_i  b_i^\dagger |i\rangle$     
with removal energies $\varepsilon_i$ and $\varepsilon_i-\omega_0$, respectively.
Here $|i \rangle$ denotes the occupied state at site ${\bf R}_i$
and the removal self-energy is given by 
$\Sigma^-(\omega)=g_i^2 \langle n_i\rangle /(\omega-\varepsilon_i+\omega_0)$.
The above variant '2' thus has an asymmetry which mimics  a strong
charge modulated state  in the sense
that on sites with large (small) charge density the self-energy is peaked
at $\omega=\pm \omega_o$ ($\varepsilon_i=0$).

On the other hand the variant '1' can be motivated from the weakly charge
modulated limit where the intensity of the bosonic satellites
at $\pm \omega_0$ is approximately the same at each site.

In cuprates evidence for inhomogeneity in the frequencies and coupling
constants of vibrational modes has been found by STM experiments
in Ref. \cite{lee2006}. According to our previous discussion 
a situation where the 
charge carriers are inhomogeneously and weakly coupled to these modes,
would correspond to variant 1. On the other hand, a stronger inhomogeneous
coupling, which would also result in a stronger electronic inhomogeneity
and thus a stronger variation of the local phonon addition and removal spectra,
could be formally captured by variant 2.

In any case this phenomenological theory reproduces our initial scenario
of low energy protected quasiparticles but emerging spectral
properties of CO scattering at large energies. In contrast
to the Kampf-Schrieffer approach which is based on homogeneous 
ground states the present theory explicitely describes
systems with broken symmetry. 

\begin{figure*}[htb]
\includegraphics[width=18cm,clip=true]{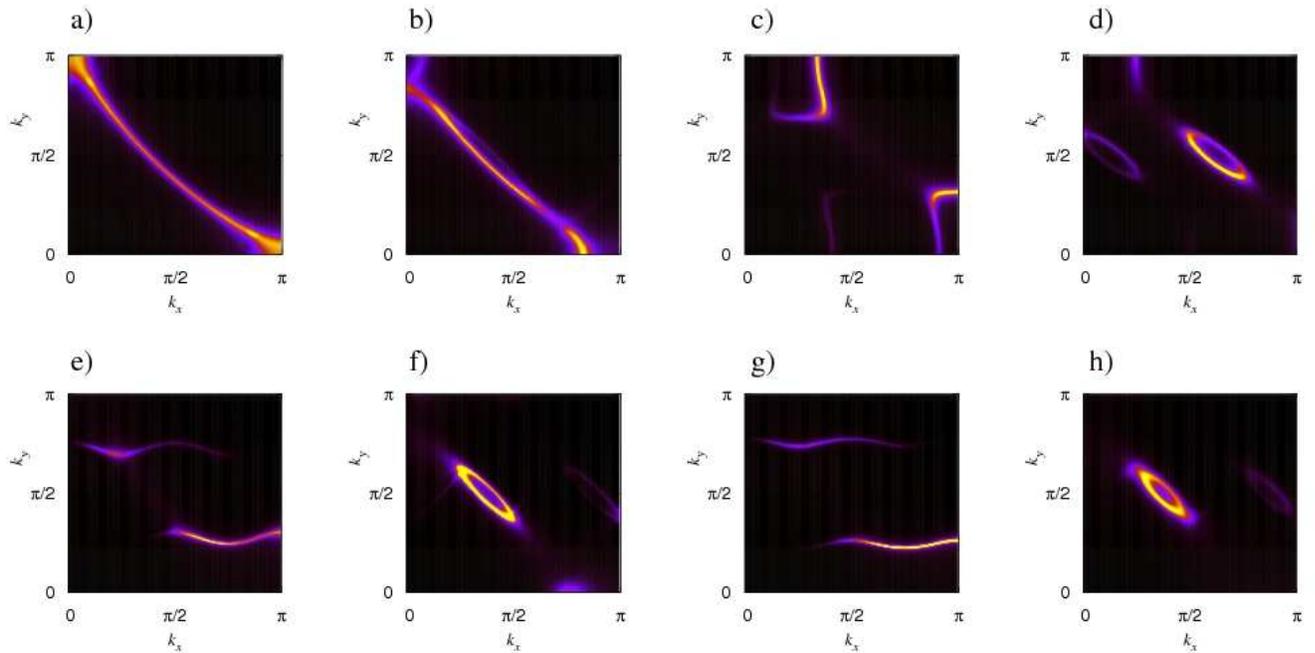}
\caption{
Constant energy scans of the spectral function for dynamic (a-f) and
static (g,h) stripes
at (a,g) $0 meV$, (b) $5 meV$, (c) $80 meV$, (d) $100 meV$, (e) $150 meV$,
(f) $250 meV$, and
(h) $200 meV$ below $E_F$. 
Window of integration: $\delta\omega=\pm 1 meV$. The energy scale
in the two-pole Ansatz for variants 1,2 is $\omega_0=50 meV$.
Figs. (a) and (b) look the same for both variants 1 and 2. The panels
(c) and (d) are obtained implementing variant 1, while panels (e) and (f)
are obtained within variant 2.
}
\label{figstr}
\end{figure*}
In Ref. \cite{sei09} we have used the approach corresponding to 'variant 1' 
in order to
investigate the bias dependence of the LDOS in connection
with the STM experiments in Ref. \cite{kohsaka08}. 
Besides the fact that the theory can reconcile                                  
the simultaneous existence of low energy Bogoljubov quasiparticles            
and high energy electronic order, it can further account for the
CO specific  contrast reversal in the STM spectra between positive and
negative bias \cite{ma08}, where the energy scale for the modulation 
of the local density
of states is essentially determined by the pairing gap.
Furthermore, from Eq. (\ref{eq:self}) it turns out that the
scattering rate (i.e. the $\qvec=0$ contribution to the self-energy) 
is determined by the sum of              
amplitudes of the electronic inhomogeneity $\sim v^2_0$.                      
Such an intrinsic relation between electronic inhomogeneity and               
inelastic scattering rate has been recently revealed by STM experiments       
on Bi2212 materials \cite{alldredge08} where it has been shown                
that the LDOS spectra can be parametrized based on                            
a model with SC $d$-wave order supplemented by                                  
an energy dependent scattering rate $\Gamma^{LDOS}_\omega=\alpha \omega$.     
The parameter $\alpha$ varies spatially and in the regions with               
pronounced charge order acquires values up to $\alpha \approx 0.4$. 
In Ref. \cite{sei09} we have used this experimental finding to approximate
the function $f(\omega)$ by a marginal-Fermi liquid type self-energy
\cite{varma91}.

Here we supplement these investigations by an alternative choice for
$f(\omega)$ which is motivated from the observation of kink structures
at some energy scale in the LDOS of STM experiments 
on cuprate superconductors that separate
homogeneous from inhomogeneous electronic states \cite{alldredge08}.
Such kinks arise via a peak in the imaginary part of the self-energy
which suggests the implementation of Eq. (\ref{eq2p}) for the 
frequency dependent spectrum $f(\omega)$.
For the real space modulation we take an array of  bond-centered stripes
separated by four lattice constants
(cf. Fig.  \ref{figlevel}). Thus the present model corresponds to
combined dynamical charge and spin order. Our aim here is  
 to establish a connection with the
ARPES experiments of Refs. \cite{zho99,zho01}.
In fact, motivated by our one-dimensional example at the beginning of
this section we investigate the problem whether our present approach can 
capture the experimentally observed dichotomy in the FS
structure and reproduce the characteristic features of straight
FS segments at higher binding energies.
It should be noted that ARPES experiments suggest that the 
true self-energy should be a combination of both parts, i.e. some bosonic
peaky feature \cite{lanzara1,GWEON,garcia} and a MFL part 
\cite{valla,johnson01} (cf. also Fig. \ref{fig-disp}). Here our 
aim is to implement these frequency structures in a scheme which
allows for symmetry broken solutions. Since the MFL part has
already been investigated in Ref. \cite{sei09} in the following we focus
on the two pole Ansatz Eq. (\ref{eq2p}) and Eqs. (\ref{eq:as2}, \ref{eq:s2}), 
respectively. 

In order to implement the stripe modulation in our approach we
calculate the corresponding local chemical potential variation 
$\lambda_i (>0)$ (sketched in Fig. \ref{figlevel})
within an unrestricted Gutzwiller approximation \cite{goe98}. 
The $\lambda_i$ can be used to construct the self-energies 
Eq. (\ref{eq:self}) for dynamical stripes in terms of variants 1,2
in the following way.  
For variant 1 we make the
correspondence $\lambda_i = \alpha v_i^2$, $\Delta_i=0$ 
with some scaling parameter $\alpha$. 
For the static solution
the correspondence simply reads $v_i^2=0$, $\Delta_i=\lambda_i$.
In case of variant 2 the two-pole Ansatz is extended as follows
\begin{eqnarray}
f_i(\omega)&=&\left\lbrack \frac{\alpha_i}{\omega-\omega_0}
+\frac{\beta_i}{\omega+\omega_0}\right\rbrack  \\
\Delta_i&=& \frac{\alpha_i-\beta_i}{\omega_0} \label{eqs3}
\end{eqnarray}
with $\alpha_i+\beta_i=1$ and $v_i^2\equiv v^2$ (cf. Eq. (\ref{eq:self})). 
Thus at each site an asymmetry is introduced in
the spectral distribution which is taken to be proportional to the 
charge modulation.
Denoting by $\overline{\lambda}\equiv \lambda_{\qvec=0}$ the average local 
chemical potential the weights of the poles are implemented as
$\alpha_i=1/2 [1-tanh(1-\lambda_i/\overline{\lambda})]$.
As in case of the commensurate example, the static component Eq. (\ref{eqs3})
is necessary to guarantee the vanishing of the scattering at low energy
($f_i(\omega=0)=0$). In addition it leads to a non-vanishing off-diagonal
Green's function at large frequencies.

Notice that for the sake of simplicity, for both variants 1,2, 
our phenomenological form of $f(\omega)$ has 
the same frequency structure for both dynamical charge- and spin scattering.

Finally, the underlying bare dispersion is 
\begin{equation}
\varepsilon_k = -2t \left[ \cos(k_x)+\cos(k_y)\right] -4t'\cos(k_x)\cos(k_y)
-\mu
\end{equation}
with $t=250meV$ and $t'/t=-0.2$ as appropriate for lanthanum cuprates 
\cite{pav01}. The chemical potential is adjusted to yield a doping
$n=0.12$ for all following results.

Panels (g,h) display cuts of the spectral function for the 
static stripe solution at $E_F$ (panel g) and $200 meV$ below $E_F$
(panel h). Naturally the one-dimensional nature of the ground state 
leads to almost straight segments in the 'FS' where the
residual 'wiggly' structure is due to the overlap of the wave-functions
between adjacent stripes. For the chosen doping $n=0.12$ and stripe separation
($4$ lattice constants), stripes are almost half-filled so that the main segments
appear at momentum $k_y=\pm \pi/4$ and weaker higher harmonic structures at  
$k_y=\pm 3\pi/4$. At higher binding energies the cuts of the
static solution reveal the AF order associated with the stripe solution.
Panel (h) shows the corresponding pockets around the nodal points at
a binding energy of $\omega=200 meV$.
Let us now turn to the resulting cuts for the dynamical stripe solutions
calculated within variant 1 (panels a,b,c,d) and variant 2 (panels a,b,e,f).  
For both dynamical stripe variants the scattering 
on the FS vanishes by construction and correspondingly panel (a)
in Fig. \ref{figstr} just displays
the bare electronic structure at $\omega=E_F$.
Below the binding energy scale $\omega_0=50 meV$ the dispersion for both variants
is gradually renormalized and for the present parameters 
changes to a large electron like FS already for $5 meV$ below E$_F$
(cf. panel (b) which is the same for variants 1,2).
Can we reproduce the formation of 'stripe segments' within the 
dynamical stripe variants at binding energies beyond the energy scale 
$\omega_0$ ?
Panels (c,d) show the high-energy  cuts of the spectral function for
variant 1. As discussed before, the scattering induced by the corresponding 
off-diagonal self-energy $\Sigma_{\Qvec \ne 0}$ vanishes for $\omega \gg \omega_0$
and thus can lead to a reconstruction of electronic states only in 
a restricted frequency range above $\omega_0$. One can in fact observe 
at $\omega=80 meV$ (panel (c)) the appearance of a small 'segment' feature 
at $k_y=\pi/4$ but the scattering is not sufficient to 'gap' completely
the electronic structure at $k_y=\pi$.
Nevertheless at slightly larger binding energies ($\omega=100 meV$, panel d)
the spectral distribution shows the characteristic pocket feature reminiscent
of the associated AF order.
Panels (e,f) show the analogous cuts for the implementation of dynamical
stripes within variant 2. Due to the fact that in this case the high energy 
scattering completely resembles that of the static solution one now
indeed observes in panel (e) the same segment structure as for the
FS of static stripes (panel g), but now at large binding
energies ($\omega=150 meV$). Moreover the spectral function changes
to the characteristic AF pocket structure at even higher energies
($\omega=250 meV$, panel (f)), again similar to the static solution reported in 
panel (h). Therefore, quite remarkably, assuming a site-dependent
fluctuation spectrum, the spectral function has the desired characteristics:
A uniform FL at low energy and spectral features typical of stripes at high energy.

\section{Conclusions}
In the present paper we have investigated the consequences of
dynamical charge (and spin) order on the spectral properties
of cuprate superconductors. We have introduced two phenomenological schemes
which can account for the dichotomy between low energy 'unperturbed'
quasiparticles and high energy charge (and spin) order as
indicated by STM \cite{kohsaka08} and ARPES \cite{zho99,zho01}
experiments.
Our first approach is a generalization of the Kampf-Schrieffer method
\cite{KAMPF} originally introduced to deal with fluctuating AF order in the
vicinity of a magnetic quantum phase transition.
At large binding 
energies one obtains a diagonal Green's function which has the same
analytic structure as that of the broken symmetry state. 
However, since the off-diagonal GF vanishes the homogeneity of the 
ground state is preserved.
By applying the Kampf-Schrieffer model to dynamical incommensurate  CO 
fluctuations in a $d$-wave superconductor we have shown that the associated 
scattering can reduce the weight of the Bogoliubov coherence peaks in the 
vicinity
of the antinodal points in agreement with ARPES data on Bi2201 compounds
\cite{kondo09}. 

We also considered another phenomenological approach,
which is based on the mapping to an auxiliary hamiltonian, 
appropriate for systems with spontaneous breaking of symmetry. 
This allows for the construction
of wave-functions which are homogeneous for energies close to $E_F$,
but manifest the symmetry-broken nature above some energy scale.
We have seen that the most promising way for such a construction is
via a coupling of the charge carriers to a spatially varying fluctuation
spectrum which has been taken to have the symmetry of charge- and spin
stripes. In this way we obtain the quite non trivial
result for the spectral intensity reported in panels (a),(b),(e),(f) of Fig. \ref{figstr},
where, upon increasing energy, a large LDA FS gradually evolves into a ``segment-like''
energy profile typical of well-formed stripes and, at even larger energies, into hole
pockets typical of a antiferromagnetically ordered state.
This result clearly captures the physical 
idea of slowly fluctuating charge collective modes, and more fastly fluctuating spin
degrees of freedom. The simple analytic (peak-like) form of the phenomenological 
self-energy also provides clear hints and constraints for the outcomes of forthcoming
microscopic theories. 

\vspace{1 truecm}
{\bf Acknowledgments} M.G. and J.L. acknowledge financial support from the MIUR project
PRIN07 prot. $2007FW3MJX_003$. 
M.G and G. S. acknowledge support from the Vigoni foundation. The work of
G.S. has been partially funded by the Deutsche Forschungsgemeinschaft.


\begin{thebibliography}{10}
\bibitem{note} Note that $\Sigma_{Q=\pi}$ is an off-diagonal self-energy
as discussed in Sec. \ref{sec:formal}.
\end{thebibliography}
\end{document}